\documentclass[a4paper,11pt]{article}

\usepackage{jheppub}

\usepackage{subfig}
\usepackage{amsmath}
\usepackage{amssymb}
\usepackage{latexsym}
\usepackage{wasysym}
\usepackage{pstricks}

\newcommand{\msbar}{\overline{\mbox{MS}}}

\newcommand{\ice}[1]{\relax}

\newcommand{\ed}{\end{document}}

\newcommand{\ep}{\epsilon}
\newcommand{\beq}{\begin{equation}}
\newcommand{\eeq}{\end{equation}}
\newcommand{\bea}{\begin{eqnarray}}
\newcommand{\eea}{\end{eqnarray}}

\newcommand{\ba}{\begin{array}}
\newcommand{\ea}{\end{array}}

\newcommand{\als}{\boldmath{\alpha_s}}

\newcommand{\g}{\gamma}
\newcommand{\be}{\beta}
\newcommand{\bc}{\begin{center}}
\newcommand{\ec}{\end{center}}

\newcommand{\re}[1]{(\ref{#1})}

\newcommand{\ovl}[1]{\overline{#1}}

\catcode`\@=11
\def\slash{\mathpalette\make@slash}
\def\make@slash#1#2{\setbox\z@\hbox{$#1#2$}%
  \hbox to 0pt{\hss$#1/$\hss\kern-\wd0}\box0}
\catcode`\@=12 

\def\bbuildrel#1_#2^#3%
{\mathrel{\mathop{\kern 0pt#1}\limits_{#2}^{#3}}}

\newcommand{\sbz}{  }

\newcommand{\as}{a_s}

\newcommand{\MSbar}{\ensuremath{\overline{\text{MS}}}}

\ice{
SFB/CPP-14-16
Title: Quark Mass and Field Anomalous Dimensions to ${\cal O}(\alpha_s^5)= $ 
}
\title{
\mbox{
\boldmath
\!\!\!\!\!\!\!\!
Quark Mass and Field  Anomalous Dimensions 
to ${\cal O}(\alpha_s^5)$
}
}
\author[a]{P.~A.~Baikov,}
\author[b]{K. G. Chetyrkin,}
\author[b]{J.~H.~K\"uhn,}

\affiliation[a]{
Skobeltsyn Institute of Nuclear Physics, Lomonosov Moscow State University, 
1(2), Leninskie gory, Moscow 119234, Russian Federation
        }
\affiliation[b]{Institut f\"ur Theoretische Teilchenphysik, Karlsruhe
  Institute of Technology (KIT), Wolfgang-Gaede-Stra\ss{}e 1, 726128 Karlsruhe, Germany}

\emailAdd{baikov@theory.sinp.msu.ru}
\emailAdd{Konstantin.Chetyrkin@kit.edu}
\emailAdd{johann.kuehn@kit.edu}

\abstract{

\rput(8.2,14){\hfill \hspace{6cm}SFB/CPP-14-16, TTP14-007}

We present  the  results of  the first complete  analytic calculation of the 
quark mass and field anomalous dimensions to ${\cal O}(\alpha_s^5)$ in QCD. }


\begin{document}

\maketitle

\section{Introduction \label{sec:intro}}

The quark masses depend on  a  renormalization scale.  The dependence
is usually referred to as ``running'' and is governed by
the quark mass anomalous dimension, $\g_m$, defined as:
\begin{equation}
\mu^2\frac{d}{d\mu^2} {m}|{{}_{{g^0},
 m^0 }}
 = {m} \gamma_m(a_s) \equiv
-{m}\sum_{i\geq0}\gamma_{{i}}
\,
a_s^{i+1}
{},
\label{anom-mass-def}
\end{equation}
where $a_s = \alpha_s/\pi= g^2/(4\pi^2)$, $g$ is the renormalized
strong coupling constant and $\mu$ is the normalization scale in the
customarily used $\msbar$ renormalization scheme. Up to and including
four loop level the anomalous dimension is known since long
\cite{Tarrach:1980up,Tarasov:1982gk,Larin:1993tq,Chetyrkin:1997dh,Vermaseren:1997fq}.
In this  paper we will describe the results of calculation of $\g_m$ and a related quantity --- the quark field anomalous dimension --- in the five-loop  order.

The evaluation of the quark mass anomalous dimension with five-loop accuracy has 
important implications. The Higgs boson decay rate into charm and
bottom quarks is proportional to the square of  the respective quark mass
at the scale of $m_H$ and the uncertainty from the presently unknown
5-loop terms in the running of the quark mass  is of order
$10^{-3}$.  This is comparable to  the precision advocated
for experiments e.g. at TLEP \cite{Gomez-Ceballos:2013zzn}. Similarly, the  issue of
Yukawa unification is affected  by precise  predictions for  the anomalous quark mass dimension.

The paper is organized as follows. The next section deals with the
overall set-up of the calculations.  Then we present our results
(Section 3), and a brief discussion (Section 4) as well as a couple of
selected applications (Section 5). Our short conclusions are given
Section 6.

\section{Technical preliminaries}
To calculate  $\gamma_m$ one needs to 
find the so-called quark mass renormalization constant, 
$Z_{m}$, which is defined as the ratio of the bare and renormalized
quark masses, viz.  
\beq
Z_m = \frac{m^0}{m} = 1  +
\sum_{i,j}^{0<j\leq i}\left(Z_{{m}}\right)_{ij}
\frac{ \as^i   }{\epsilon^{{j}}}
\label{Zm}
{}.
\eeq
Within the
\MSbar\  scheme \cite{tHooft:1972fi,Bardeen:1978yd}
the coefficients $\left(Z_{{m}}\right)_{ij}$ are just numbers
\cite{Collins:1974da}; $\epsilon \equiv 2 - D/2 $ and $D$ stands for
the space-time dimension. Combining eqs.~(\ref{anom-mass-def},\ref{Zm}) and
using the RG-invariance of of $m^0$, one arrives at the following
formula for $\g_m$:
\beq
\gamma_m = \sum_{ i \ge 0}  (Z_m)_{i1} \,i\,  a_s^i
{}.
\eeq
To find $Z_m$  one should    compute the vector and
scalar parts of the quark self-energy 
$\Sigma_V(p^2)$ and $\Sigma_S(p^2)$.
In our convention, the bare quark propagator is proportional to 
$\left[\not\!p\left(1+\Sigma_V^0(p^2)\right)
-m_q^0\left(1-\Sigma_S^0(p^2)\right)\right]^{-1}
$. Requiring
the finiteness of the renormalized quark propagator 
and keeping only massless and  terms linear in $m_q$, one  arrives
at  the following recursive   equations  to find $Z_m$ 
\beq
Z_m Z_2 = 1 +  K_\ep
\left\{ 
Z_m Z_2 \Sigma_S^0(p^2)
\right\}
, \ \ \ 
Z_2 =1 - K_\ep
\left\{ 
Z_2  \Sigma_V^0(p^2)
\right\}
\label{Z2Zm}
{},
\eeq
where $K_\epsilon \left\{f(\epsilon)\right\}$ stands for the singular part of the
Laurent expansion of $f(\epsilon)$ in $\epsilon$ near $\epsilon=0$ and
$Z_2$ is the quark wave function renormalization constant.
Eqs. (\ref{Z2Zm}) express $Z_m$  through massless propagator-type 
(that is dependent on  one external momentum only) 
Feynman integrals (FI), denoted  as  {\em p-integrals} below.

Eqs. (\ref{Z2Zm}) require the calculation of a large  number\footnote{
We have used  QGRAF \cite{Nogueira:1991ex} to  produce around $10^5$ 
FI's  contributing to  the quark self-energy at ${\cal O}(\alpha_s^5)$.}
the {\em five}-loop p-integrals to find $Z_m$ and $Z_2$  to
${\cal O}(\alpha_s^5)$.  

At present there exists no direct way to analytically evaluate five-loop
p-integrals.  However, according to \re{Zm} for a given five-loop
p-integral we need to know only its {\em pole} part in $\ep$ in the
limit of $\ep \to 0$. A proper use of this fact can significantly
simplify our task. The corresponding method---so-called Infrared
Rearrangement (IRR)---first suggested in \cite{Vladimirov:1979zm} and
elaborated further in \cite{Kazakov:1979ik,Chetyrkin:1980pr,Tarasov:1980au} 
allows to effectively decrease number of
loops to be computed by one\footnote{With the price that resulting
one-loop-less p-integrals should be evaluated up to and {\em
including} their constant part in the small $\ep$-expansion.}.  In its
initial version IRR was not really universal; it was not applicable in
some (though rather rare) cases of complicated FI's.  The problem was
solved by  elaborating a special technique of subtraction of IR
divergences --- the $R^*$-operation
\cite{Chetyrkin:1984xa,Chetyrkin:1996ez}. This technique succeeds in
expressing the UV counterterm of every L-loop Feynman integral in
terms of divergent and finite parts of some (L-1)-loop massless
propagators.

In our case $L=5$ and, using IRR, one arrives at at around $10^5$
four-loop p-integrals. These can, subsequently, be reduced to 28
four-loop masterp-integrals, which are known analytically, including
their finite parts, from \cite{Baikov:2010hf,Lee:2011jt} as well as
numerically from \cite{Smirnov:2010hd}.

We need, thus,  to   compute around $10^5$   p-integrals.
Their singular parts, in turn, can be algebraically reduced to only 28
master 4-loop p-integrals.  The reduction is based on evaluating
sufficiently many terms of the $1/D$ expansion \cite{Baikov:2005nv} of
the corresponding coefficient functions \cite{Baikov:1996rk}.

All our calculations   have  been performed
on a SGI ALTIX 24-node IB-interconnected cluster of eight-cores Xeon
computers using  parallel  MPI-based \cite{Tentyukov:2004hz} as well as thread-based 
\cite{Tentyukov:2007mu} versions  of FORM
\cite{Vermaseren:2000nd}.   

\section{Results}

Our result for the  anomalous dimension 
\[\gamma_{m} = -\sum_{i=0}^{\infty} \ (\gamma_{m})_i \, a_s^{i+1}\]
reads:
\beq
(\g_m)0 = 1,
\ \ \ \
(\g_m)1 = \frac{1}{16}
\left\{ \rule{0cm}{6mm} \right.
 \frac{202}{3}
{+} \,n_f 
\left[
-\frac{20}{9}\right]
\left. \rule{0cm}{6mm} \right\}
{},
\eeq
\beq
(\g_m)2 = \frac{1}{64} 
\left\{ \rule{0cm}{6mm} \right.
 1249
{+} \,n_f 
\left[
-\frac{2216}{27} 
-\frac{160}{3}  \,\zeta_3
\right]
{+} \, n_f^2
\left[
-\frac{140}{81}\right]
\left. \rule{0cm}{6mm} \right\}
{},
\eeq
\begin{eqnarray}
\lefteqn{(\g_m)3= \frac{1}{256} 
\left\{ \rule{0cm}{6mm} \right.
\frac{4603055}{162} 
+\frac{135680}{27}  \,\zeta_3
-8800  \,\zeta_5
}
\nonumber\\
&{+}& \,n_f 
\left[
-\frac{91723}{27} 
-\frac{34192}{9}  \,\zeta_3
+880  \,\zeta_4
+\frac{18400}{9}  \,\zeta_5
\right]
\nonumber\\
&{+}& \, n_f^2
\left[
\frac{5242}{243} 
+\frac{800}{9}  \,\zeta_3
-\frac{160}{3}  \,\zeta_4
\right]
{+} \, n_f^3
\left[
-\frac{332}{243} 
+\frac{64}{27}  \,\zeta_3
\right]
\left. \rule{0cm}{6mm} \right\}
{}.
\end{eqnarray}
\begin{eqnarray}
(\gamma_m)_{4} &=& 
\nonumber
\frac{1}{4^5} \Biggl\{
\frac{99512327}{162} 
 + \frac{46402466}{243}  \sbz \zeta_{3}
 + 96800  \,\zeta_3^2
 - \frac{698126}{9}  \sbz \zeta_{4}
\nonumber\\
&{}& \hspace{1.2cm}
 - \frac{231757160}{243}  \sbz \zeta_{5}
 + 242000  \,\zeta_{6}
 + 412720  \,\zeta_{7}
\nonumber\\
&{+}& \, n_f 
\left[
-\frac{150736283}{1458} 
 - \frac{12538016}{81}  \sbz \zeta_{3}
 - \frac{75680}{9}  \,\zeta_3^2
 + \frac{2038742}{27}  \sbz \zeta_{4}
\right.
\nonumber
\\
&{}& 
\left. 
\hspace{1.2cm}
 + \frac{49876180}{243}  \sbz \zeta_{5}
 - \frac{638000}{9}  \,\zeta_{6}
 - \frac{1820000}{27}  \,\zeta_{7}
\right]
\label{gm5}
\\
&{+}& \, n_f^2
\left[
\frac{1320742}{729} 
 + \frac{2010824}{243}  \sbz \zeta_{3}
 + \frac{46400}{27}  \,\zeta_3^2
 - \frac{166300}{27}  \sbz \zeta_{4}
 - \frac{264040}{81}  \sbz \zeta_{5}
 + \frac{92000}{27}  \,\zeta_{6}
\right]
\nonumber\\
&{+}& \,
\fbox{$
 n_f^3
\left[
\frac{91865}{1458} 
 + \frac{12848}{81}  \sbz \zeta_{3}
 + \frac{448}{9}  \sbz \zeta_{4}
 - \frac{5120}{27}  \sbz \zeta_{5}
\right]
+ \, n_f^4
\left[
-\frac{260}{243} 
 - \frac{320}{243}  \sbz \zeta_{3}
 + \frac{64}{27}  \sbz \zeta_{4}
\right]
$}
\Biggr\}
{}.
\nonumber
\end{eqnarray}
Here $\zeta$ is the Riemann zeta-function 
($\zeta_3 = 1.202056903\dots$, $\zeta_4 = \pi^4/90$,
$\zeta_5  = 1.036927755\dots$, 
$\zeta_6  = 1.017343062\dots$ and
$\zeta_7  = 1.008349277\dots$
).
Note that in four-loop order we exactly\footnote{This agreement can be also considered as an important check of all our setup which is completely different from the ones utilized at the four-loop calculations.}  reproduce well-known results obtained in 
\cite{Chetyrkin:1997dh,Vermaseren:1997fq}.
The  boxed terms in \re{gm5} are in full agreement with the results derived previously  on 
the basis of the $1/n_f$ method in \cite{PalanquesMestre:1983zy,Ciuchini:1999wy,Ciuchini:1999cv}.

For completeness we  present below the result for the quark field anomalous dimension 
$\gamma_{2} = -\sum_{i=0}^{\infty} \ (\gamma_{2})_i \, a_s^{i+1}$: 
\begin{eqnarray}
(\gamma_2)_{4} &=&   
\nonumber
\frac{1}{4^5} \Biggl\{
\frac{2798900231}{7776} 
+\frac{17969627}{864}  \sbz \zeta_{3}
+\frac{13214911}{648}  \,\zeta_3^2
+\frac{16730765}{864}  \sbz \zeta_{4}
-\frac{832567417}{3888}  \sbz \zeta_{5}
\\
\nonumber
&{}& \hspace{0.8cm}
+\frac{40109575}{1296}  \,\zeta_{6}
+\frac{124597529}{1728}  \,\zeta_{7}
\nonumber\\
&{+}& \, n_f 
\left[
-\frac{861347053}{11664} 
-\frac{274621439}{11664}  \sbz \zeta_{3}
+\frac{1960337}{972}  \,\zeta_3^2
+\frac{465395}{1296}  \sbz \zeta_{4}
\right.
\nonumber\\
&{}& \hspace{.8cm}
\left.
+\frac{22169149}{5832}  \sbz \zeta_{5}
+\frac{1278475}{1944}  \,\zeta_{6}
+\frac{3443909}{216}  \,\zeta_{7}
\right]
\nonumber\\
&{+}& \, n_f^2
\left[
\frac{37300355}{11664} 
+\frac{1349831}{486}  \sbz \zeta_{3}
-\frac{128}{9}  \,\zeta_3^2
-\frac{27415}{54}  \sbz \zeta_{4}
-\frac{12079}{27}  \sbz \zeta_{5}
-\frac{800}{9}  \,\zeta_{6}
-\frac{1323}{2}  \,\zeta_{7}
\right]
\nonumber\\
&{+}& \, n_f^3
\left[
-\frac{114049}{8748} 
-\frac{1396}{81}  \sbz \zeta_{3}
+\frac{208}{9}  \sbz \zeta_{4}
\right]
{+} \, n_f^4
\left[
\frac{332}{729} 
-\frac{64}{81}  \sbz \zeta_{3}
\right]
\Biggr\}
{}.
\label{g25l}
\end{eqnarray}
The above result is presented for the Feynman gauge; the coefficients
$(\g_2)_i $ with $i \le 3$ can be found in \cite{Chetyrkin:1999pq}
(for the case of a general covariant gauge and SU(N) gauge group).

\section{Discussion}

In  numerical form $\g_m$ reads
\bea
\nonumber
\g_m =  &-& a_s - a_s^2 \left(4.20833 - 0.138889 n_f\right)
\\ \nonumber
&-&
a_s^3  \left(19.5156 - 2.28412 n_f - 0.0270062 n_f^2 \right)  
\\ \nonumber 
&-&
a_s^4 \left(98.9434 - 19.1075 n_f + 0.276163 n_f^2  + 0.00579322 n_f^3 \right)
\\
&-&
a_s^5 \left(
559.7069 - 143.6864\, n_f + 7.4824\, n_f^2  + 0.1083\, n_f^3  - 0.000085359\, n_f^4
\right)
{}
\label{N[gm5qcd]}
\eea
\ice{
In[12]:= Coll[gm5/.Zrule//N, {as,nf}]//InputForm

Out[12]//InputForm= 
-1.*as + as^2*(-4.208333333333333 + 0.1388888888888889*nf) + 
 as^3*(-19.515625 + 2.284121493373736*nf + 0.02700617283950617*nf^2) + 
 as^4*(-98.9434142552029 + 19.1074619186354*nf - 0.27616255142989465*nf^2 - 
   0.005793222354358382*nf^3) + 
 as^5*(-559.706895735422 + 143.68646716196457*nf - 7.482378227164126*nf^2 - 
   0.10831832516187696*nf^3 + 0.00008535890236567112*nf^4)

}
and
 \bea
\nonumber                               
\g_m \bbuildrel{=\!=\!=}_{n_f = 3}^{}
&-& \as - 3.79167 \,\as^2  - 12.4202 \,\as^3  - 44.2629 \,\as^4  - 198.907 \,\as^5
 ,\\ \nonumber                                            
g_m \bbuildrel{=\!=\!=}_{n_f = 4}^{}
&-& \as - 3.65278 \,\as^2  - 9.94704 \,\as^3  - 27.3029 \,\as^4  - 111.59 \,\as^5
,\\  \nonumber 
\ g_m \bbuildrel{=\!=\!=}_{n_f = 5}^{}                                                      
&-& \as - 3.51389 \,\as^2  - 7.41986 \,\as^3  - 11.0343 \,\as^4  - 41.8205 \,\as^5
 ,\\      
\g_m \bbuildrel{=\!=\!=}_{n_f = 6}^{}
&-&  \as - 3.37500   \,\as^2  - 4.83867 \,\as^3  + 4.50817 \,\as^4  + 9.76016 \,\as^5
\label{gm5:nf:3-6}
{}.
\eea
Note that significant cancellations between $n_f^0$ and $n_f^1$ terms
for the values of $n_f$ around 3 or so persist also at five-loop
order.  As a result we observe  a moderate growth of the series
in $\as$ appearing in the quark mass anomalous dimension at various
values of active quark flavours (recall that even for scales as small
as 2 GeV $\as \equiv \frac{\alpha_s}{\pi} \approx 0.1$).

Similar behavior  shows up  for $\g_2$: 
\bea
\nonumber
\g_2 =  &-& 0.33333  a_s - a_s^2 \left(-1.9583  +0.08333 \,n_f\right)
\\ \nonumber
&-&
a_s^3  \left(-10.3370  +1.0877 \,n_f -0.01157 \,n_f^{2} \right)  
\\ \nonumber 
&-&
a_s^4 \left( -53.0220  +10.1090 \,n_f -0.27703 \,n_f^{2} -0.0023 \,n_f^{3}\right)
\\
&-&
a_s^4 \left(
 -310.0700  +76.3260 \,n_f -4.6339 \,n_f^{2} +0.0085 \,n_f^{3}  +0.00048 \,n_f^{4}
\right)
{}
\label{N[gm5qcd]}
\eea

 \ice{
 -0.33333  
 -1.9583  +0.08333 \,n_f

 -10.3370  +1.0877 \,n_f -0.01157 \,n_f^{2}

 -53.0220  +10.1090 \,n_f -0.27703 \,n_f^{2} -0.0023 \,n_f^{3}

 -310.0700  +76.3260 \,n_f -4.6339 \,n_f^{2} +0.0085 \,n_f^{3}  +0.00048 \,n_f^{4}

}

and
 \bea
\nonumber                               
\g_2 \bbuildrel{=\!=\!=}_{n_f = 3}^{}
&-& 0.33333\,\as -1.7083 \,a_s^{2} -7.1779 \,a_s^{3} -25.2480 \,a_s^{4}   -122.5300 \,a_s^{5}
 ,\\ \nonumber                                            
g_2 \bbuildrel{=\!=\!=}_{n_f = 4}^{}
&-&0.33333 \,\as  -1.6250 \,a_s^{2} -6.1712 \,a_s^{3} -17.1610 \,a_s^{4}  -78.2430 \,a_s^{5}
,\\  \nonumber 
\ g_2 \bbuildrel{=\!=\!=}_{n_f = 5}^{}                                                      
&-&0.33333 \,\as -1.5417 \,a_s^{2} -5.1877 \,a_s^{3} -9.6824 \,a_s^{4} -42.9240 \,a_s^{5}
 ,\\      
\g_2 \bbuildrel{=\!=\!=}_{n_f = 6}^{}
&-& 0.33333 \,\as -1.4583 \,a_s^{2} -4.2274 \,a_s^{3} -2.8251 \,a_s^{4} -16.4710 \,a_s^{5}
\label{gm5:nf:3-6}
{}.
\eea
\ice{

 -0.33333 \,a_s 
-1.7083 \,a_s^{2} -7.1779 \,a_s^{3} -25.2480 \,a_s^{4}   -122.5300 \,a_s^{5}

 -0.33333 
\,a_s -1.6250 \,a_s^{2} -6.1712 \,a_s^{3} -17.1610 \,a_s^{4}  -78.2430 \,a_s^{5}

 -0.33333 \,a_s 
-1.5417 \,a_s^{2} -5.1877 \,a_s^{3} -9.6824 \,a_s^{4} -42.9240 \,a_s^{5}

 -0.33333 \,a_s 
-1.4583 \,a_s^{2} -4.2274 \,a_s^{3} -2.8251 \,a_s^{4} -16.4710 \,a_s^{5}

   }
It is instructive to compare our numerical result for $(\g_m)_4$
\beq 
(\g_m)_4  =  559.71 - 143.6\, n_f + 7.4824\, n_f^2  + 0.1083\, n_f^3  - 0.00008535\, n_f^4 
\label{gm4_num}
{}
\eeq
with a 15 years old prediction based on the 
``Asymptotic P\'ade Approximants'' (APAP) method \cite{Ellis:1997sb} (the boxed term  below 
was used as the  input)
\beq 
(\g_m)_4^{\rm APAP}   =  530 - 143\, n_f + 6.67\, n_f^2  + 0.037\, n_f^3   -\fbox{$ 0.00008535\,n_f^4$} 
\eeq
Unfortunately, this impressively good agreement does {\em not} survive
for fixed values of $n_f$ due to severe cancellations between different powers of $n_f$ as one can see from the Table 1.
\begin{table}
\begin{center}
  \begin{tabular}{| c | c | c|c|c| }
    \hline
    $n_f$               &  3   & 4  &  5  &     6   \\ \hline
    $(\g_m)_4^{\rm exact}$       &  198.899   &  111.579  &  41.807  &     -9.777 \\ [0.51mm]    \hline
    \rule{0mm}{5.5mm}
    $(\g_m)_4^{\rm APAP}$  \cite{Ellis:1997sb}         &  162.0   & 67.1  &  -13.7  &   -80.0 \\[1mm] 
    $(\g_m)_4^{\rm APAP}$  \cite{Elias:1998bi}         &  163.0   & 75.2  &  12.6  &   12.2 \\[1mm]  
    $(\g_m)_4^{\rm APAP}$  \cite{Kataev:2008ym}         &  164.0   & 71.6 &  -4.8  &   -64.6 \\ [1mm] 
    \hline
  \end{tabular}
\end{center}
\caption{The exact results for $(\g_m)_4$ together with the predictions made with the help of
the original APAP method and   its two somewhat  modified versions.}
\end{table}

\ice{
In[5]:= t1 = Table[ gm5APAP1[nf], {nf,3,6}]

Out[6]= {162.022, 67.0661, -13.6784, -79.9987}

In[7]:= t2 = Table[ gm5APAP2[nf], {nf,3,6}]

Out[7]= {163, 75.2, 12.6, 12.2}

In[8]:= t3 = Table[ gm5APAP3[nf], {nf,3,6}]

Out[8]= {163.948, 71.6306, -4.76338, -64.5936}
}
The solution of eq. \re{anom-mass-def} reads:
\beq
\frac{m(\mu)}{m(\mu_0)} = \frac{c(a_s(\mu))}{c(a_s(\mu_0))}, \  \ \ 
c(x) = \mathrm{exp}\Biggl\{ \int {d x'} \frac{\g_m(x'}{\beta(x')} \Biggr\} 
{},
\label{cfun:1}
\eeq
\begin{eqnarray}
c(x) = & {} & (x)^{\bar{\g_0}} \left\{ 1 + d_1 x + (d_1^2/2 + d_2) \,x^2 
          +  (d_1^3/6 + d_1 d_2 + d_3)\,  {x^3} \right.\nonumber \\
& + &        \left.(d_1^4/24 + d_1^2 d_2/2 + d_2^2/2+ d_1 d_3 + d_4)
\,{x^4} + {\cal O}(x^5) \right\}
\label{cfun:2}
{},
\end{eqnarray}

\ice{

Out[74]//InputForm= 
1 + d1*x + (d1^2/2 + d2)*x^2 + (d1^3/6 + d1*d2 + d3)*x^3 + 
 (d1^4/24 + (d1^2*d2)/2 + d2^2/2 + d1*d3 + d4)*x^4
}

\begin{flalign}
& d_1 = -\bar{\beta}_1 \, \bar{\gamma}_0 + \bar{\gamma}_1 {},&
\\
&d_2 = \bar{\beta}_1^2 \, \bar{\gamma}_0/2 - \bar{\beta}_2 \, \bar{\gamma}_0/2 - \bar{\beta}_1 \, \bar{\gamma}_1/2 + \bar{\gamma}_2/2
{}, &
\\
&d_3 = -\bar{\beta}_1^3 \, \bar{\gamma}_0/3 + 2 \, \bar{\beta}_1 \, \bar{\beta}_2 \, \bar{\gamma}_0/3 - \bar{\beta}_3 \, \bar{\gamma}_0/3 + \bar{\beta}_1^2 \, \bar{\gamma}_1/3 
          -\bar{\beta}_2 \, \bar{\gamma}_1/3 - \bar{\beta}_1 \, \bar{\gamma}_2/3 + \bar{\gamma}_3/3
{},
&
\\
&d_4 = \bar{\beta}_1^4 \, \bar{\gamma}_0/4 - 3 \, \bar{\beta}_1^2 \, \bar{\beta}_2 \, \bar{\gamma}_0/4 + \bar{\beta}_2^2 \, \bar{\gamma}_0/4 + \bar{\beta}_1 \, \bar{\beta}_3 \, \bar{\gamma}_0/2 
    -  \bar{\beta}_4 \, \bar{\gamma}_0/4 - \bar{\beta}_1^3 \, \bar{\gamma}_1/4 
 &
\nonumber
   \\ 
   & \hspace{9mm} + \bar{\beta}_1 \,  \bar{\beta}_2 \, \bar{\gamma}_1/2 - \bar{\beta}_3 \, \bar{\gamma}_1/4 +   \bar{\beta}_1^2 \, \bar{\gamma}_2/4 - \bar{\beta}_2 \, \bar{\gamma}_2/4 
    - \bar{\beta}_1 \, \bar{\gamma}_3/4 + \bar{\gamma}_4/4 {}.
\label{d4}
&
\end{flalign}
Here 
$\bar{\g_i} = (\g_m)_i/\beta_0$, $\bar{\beta}_i = \beta_i/\beta_0$   
 and 
\[\beta(a_s)  =-\sum_{i \ge 0} \,\beta_i \, a_s^{i+2} = -\beta_0 \left\{\sum_{i \ge 0} \,\bar{\beta_i}\,  a_s^{i+2}
\right\}
\]
is the QCD $\beta$-function. Unfortunately, the coefficient $d_4$ in
eq.~\re{d4} does depend on the yet unknown {\em five-loop} coefficient
$\beta_4$ (up to four loops the $\beta$-function is known from
\cite{Gross:1973id,Politzer:1973fx,Caswell:1974gg,Jones:1974mm,Egorian:1978zx,Tarasov:1980au,Larin:1993tp,vanRitbergen:1997va,Czakon:2004bu}).

Numerically, the $c$-function  reads:
\[
c(x)\bbuildrel{=\!=\!=}_{n_f = 3}^{}  x^{4/9}\,c_s(x), \
c(x)\bbuildrel{=\!=\!=}_{n_f = 4}^{}  x^{12/25}\,c_c(x),\ 
c(x)\bbuildrel{=\!=\!=}_{n_f = 5}^{}  x^{12/23}\,c_b(x), \
c(x)\bbuildrel{=\!=\!=}_{n_f = 6}^{}  x^{4/7}\,c_t(x)
{},
\]
with
\bea
c_s(x) &=& 1  + 0.8950 \,x + 1.3714 \,x^2  + 1.9517 \,x^3 + (15.6982-  0.11111\,\bar{\beta_4}) \,x^4    ,   
\nonumber
\\
c_c(x) &=& 1 + 1.0141 \,x + 1.3892 \,x^2  + 1.0905 \,x^3  + ( 9.1104  -  0.12000  \,\bar{\beta_4}) \,x^4     ,
\nonumber
\\
c_b(x) &=& 1+ 1.1755 \,x + 1.5007 \,x^2  + 0.17248 \,x^3  + (  2.69277 -  0.13046  \,\bar{\beta_4}) \,x^4 ,
\nonumber
\\
c_t(x) &=& 1+  1.3980 \,x + 1.7935 \,x^2  - 0.68343 \,x^3  + ( - 3.5130 -    0.14286 \,\bar{\beta_4}) \,x^4   
\label{cfunctions}
{}.
\eea

\section{Applications}
\subsection{ RGI mass}

Eq. \re{cfun:1} naturally leads to an important concept: the  RGI mass
\beq
 {m}^{\rm RGI} \equiv m(\mu_0)/{c(a_s(\mu_0))}
{}, 
\eeq
which is often used in  the context  of lattice calculations. 
The mass is $\mu$ and 
{\em scheme} independent; in  {\em any} (mass-independent) scheme
\[ \lim_{\mu \to \infty} a_s(\mu)^{-\bar{\g}_0}\, \, m(\mu) = {m}^{\rm RGI}
{}.
\]
The function $c_s(x)$ is used, e.g, by the {\bf ALPHA} lattice
collaboration to find the $\ovl{\mbox{MS}}$ mass of the strange quark
at a lower scale, say, $m_s(2 \ \mbox{GeV})$ from the $m_s^{\rm RGI}$ mass
determined from lattice simulations (see, e.g. \cite{DellaMorte:2005kg}). 
For example, setting $a_s(\mu = 2\, \mbox{GeV}) = \frac{\large\alpha_s(\mu)}{ \pi}
= 0.1$, we arrive at ($h$ counts loops):
\bea
m_s(2 \,\mbox{GeV}) &=&  m_s^{\rm RGI} \left(a_s(2\, \mbox{GeV}) \right)^{\frac{4}{9}} 
\, \Bigl(
1  + 0.0895\, h^2 + 0.0137\, h^3  + 0.00195\, h^4 
\nonumber
\\
&{}& \
\hspace{4cm}
 + (0.00157   - 0.000011 \,\ovl{\be}_4)\, h^5
\Bigr)
\label{cs2GeV}
\eea
In  order to have an idea of  effects due the  five-loop term 
in \re{cs2GeV} one should make a guess about $\bar{\beta_4}$. By inspecting
lower orders in 
 \[
\beta(n_f=3) = -\left(\frac{4}{9}\right) \,\Bigl(
\as + 1.777\, \as^2 + 4.4711\, \as^3 + 20.990\, \as^4 + \bar{\beta}_4\, \as^5
\Bigr)
\]
one can assume a natural estimate of  $\ovl{\beta}_4$ as laying in the interval $50-100$.
With this  choice we conclude  that the (apparent) convergence of
the above series is quite good even at a rather small energy scale of  2 GeV.

On the other hand, the authors of \cite{Elias:1998bi} estimate 
$\bar{\beta}_4$ in the $n_f=3$
QCD as large as -850! With such  a huge and negative  value of $\bar{\beta}_4$ the 
five loop term in \re{cs2GeV} would amount to $0.01092$ and, thus,  would significantly exceed
the four-loop contribution (0.00195).


\subsection{Higgs decay into quarks}

The decay width of the Higgs boson into  a
pair of quarks can be written in the form
\beq
\Gamma(H \to \bar{f}f )
=\frac{G_F\,M_H}{4\sqrt{2}\pi}\,
m_f^2(\mu) \,  R^S (s = M_H^2,\mu)
\label{H2ff}
\eeq
where $\mu$ is the
normalization scale and
$R^S$ is the spectral density of the scalar correlator,  known to
$\alpha_s^4$ from  \cite{Baikov:2005rw}
\bea
\hspace{-1cm}R^{S}(s = M_H^2,\mu=M_H) &=& 1 + 5.667\,  a_s+ 29.147 \, a_s^2  +
  41.758 \,a_s^3 \,  {- 825.7}\,a_s^4
\nonumber
\\
&=&
1 + 0.2041  + 0.0379  + 0.0020  {-0.00140}
\label{RS_as4_nl5}
\eea
where we set
$a_s = \alpha_s/\pi= 0.0360$ (for the Higgs mass value $M_H = 125$  GeV and $\alpha_s(M_Z) = 0.118$).

Expression \re{H2ff} depends on two phenomenological parameters,
namely, $\alpha_s(M_H)$ and the quark running mass $m_q$.  In what
follows we consider, for definiteness, the dominant decay mode $H
\to {\bar b}{b}$.  To avoid the appearance of large logarithms of the type
$\ln \mu^2/M_H^2$ the parameter $\mu$ is customarily chosen to be
around $M_H$.  However, the starting value of $m_b$ is usually
determined at a much smaller scale (typically around 5-10 GeV
\cite{Chetyrkin:2009fv}). The evolution of $m_b(\mu)$ from a lower
scale to $\mu = M_h$ is described by a corresponding RG equation which
is completely fixed by the quark mass anomalous dimension
$\gamma(\als)$ and the QCD beta function $\beta(\als)$ (for QCD with
$n_f=5$).  In order to match the ${\cal O}(\als^4)$ accuracy of
\re{RS_as4_nl5} one should know {\em both} RG functions $\beta$ and
$\gamma_m $ in the five-loop approximation.  Let us proceed, assuming
conservatively that $ 0 \le \, {\bar{\beta}_4}^{n_f=5} \le 200$.

The value of $m_b(\mu=M_H)$ is to be obtained with RG running from
$m_b(\mu=\,10 \, \mbox{GeV})$ and, thus, depends on $\beta$ and
$\g_m$.  Using the Mathematica package RunDec\footnote{We have
extended the package by including the five-loop effects to the running
of $\als$ and quark masses.}  \cite{Chetyrkin:2000yt} and eq. 
\re{cfunctions} we  find for  the shift from  the five-loop term
\[  
\frac{\delta m_b^2(M_H)}{m_b^2(M_H)} =
 -1.3\cdot 10^{-4}({\bar{\beta}_4 =0}) 
|-4.3 \cdot 10^{-4}({\bar{\beta}_4 =100})
| -7.3\cdot 10^{-4}({\bar{\beta}_4 =200})
\]
If we set $\mu=M_H$,  then the combined effect of ${\cal O}(\alpha_s^4$)
terms as coming from the five-loop
running and four-loop contribution to $R^S$ on 
\beq
\Gamma(H \to \bar{b} b )
=\frac{G_F\,M_H}{4\sqrt{2}\pi}\,
m_f^2(M_H) \,  R^S (s = M_H^2,M_H)
\label{Hbb2}
\eeq
is around {-2\permil} (for $\bar{\beta}_4 =100$).  This should be
contrasted to the parametric uncertainties  coming from the input
parameters $\alpha_s(M_Z) = 0.1185(6)$ \cite{Beringer:1900zz} and
$m_b(m_b)= 4.169(8)\ \mbox{GeV}$ \cite{Penin:2014zaa} which correspond
to {$\pm$ 1\permil}\ and {$\pm$ 4\permil}\ respectively.

\ice{
In[83]:= (4.169 + .008)^2/4.169^2
Out[83]= 1.00384
}
We conclude, that the ${\cal O}(\alpha_s^4)$ terms in \re{RS_as4_nl5},
\re{Hbb2}) are of no phenomenological relevancy at present. But, the
situation could be different if the project of TLEP
\cite{Gomez-Ceballos:2013zzn} is implemented. For instance, the
uncertainty in $\alpha_s(M_Z)$ could be reduced to $\pm 2\permil$ and
Higgs boson branching ratios with precisions in the permille range are
advertised.

\ice{
In[85]:= 
0.0006/0.1186
Out[85]= 0.00505902
}

\section{Conclusions}

We have analytically computed the anomalous dimensions of the quark
mass $\g_m$ and field $\g_2$ in the five loop approximation. The
self-consistent description of the quark mass evolution at five loop
requires the knowledge of the QCD $\beta$-function to the same number
of loops. The corresponding, significantly more
complicated calculation is under consideration.

K.G.C. thanks  J. Gracey and  members of the DESY-Zeuthen theory  seminar 
for  usefull discussions.  

This work was supported by
the Deutsche Forschungsgemeinschaft in the
Sonderforschungsbereich/Transregio
SFB/TR-9 ``Computational Particle Physics''. The work of P.~Baikov was 
supported in part by the Russian Ministry of Education and Science
under grant NSh-3042.2014.2.

\providecommand{\href}[2]{#2}\begingroup\raggedright\endgroup

\ed

\bibliographystyle{JHEP}

\bibliography{lit,gm5,higgs_as5,dim_reg,JJ,QGRAF_EXP,chet09,chet,steinhauser,baikov,asmirnov,smirnov,vladimirov,vermaseren,%
surguladze,laporta,gorishnii,tarasov,bierenbaum,kataev,czakon,kazakov,david_dirk_qQED,other_masters,%
sector_decom,lee,kotikov_before1995,remiddi_1997-2000,broadhurst,gracey,velizhanin,%
DIS,DIS2,DIS4,acat,penin}

\ed